\documentclass[
]{ceurart}

\usepackage[capitalize,noabbrev]{cleveref}
\usepackage{listings}
\usepackage{subcaption}
\usepackage{mathtools}

\usepackage{xr}
\graphicspath{{Figures/}}

\begin{document}

\copyrightyear{2026}
\copyrightclause{Copyright for this paper by its authors.
  Use permitted under Creative Commons License Attribution 4.0
  International (CC BY 4.0).}

\conference{AI CHAOS'26: Workshop Series on the Challenges for Human Oversight of AI Systems}

\title{Intelligent support for Human Oversight: Integrating Reinforcement Learning with Gaze Simulation to Personalize Highlighting}

\author[1]{Thorsten Kl\"o\ss ner}[%
email=kloessner@cs.uni-saarland.de,
]
\cormark[1]

\address[1]{Saarland University, Saarbr\"ucken, Germany}

\author[1]{João Belo}[%
email=jbelo@cs.uni-saarland.de,
]

\author[1]{Zekun Wu}[%
email=wuzekun@cs.uni-saarland.de,
]

\author[1]{J\"org Hoffmann}[%
email=hoffmann@cs.uni-saarland.de,
]

\author[1]{Anna Maria Feit}[%
email=feit@cs.uni-saarland.de,
]

\cortext[1]{Corresponding author.}

\begin{abstract}
Interfaces for human oversight must effectively support users’ situation awareness under time-critical conditions. We explore reinforcement learning (RL)–based UI adaptation to personalize alerting strategies that balance the benefits of highlighting critical events against the cognitive costs of interruptions. To enable learning without real-world deployment, we integrate models of users’ gaze behavior to simulate attentional dynamics during monitoring. Using a delivery-drone oversight scenario, we present initial results suggesting that RL-based highlighting can outperform static, rule-based approaches and discuss challenges of intelligent oversight support.
\end{abstract}


\maketitle

\section{Introduction}

As AI-driven systems gain higher degrees of autonomy, there is an increasing call for human involvement to \emph{monitor} the behavior of autonomous systems (e.g. of autonomous vehicles) and \emph{intervene} to avoid or resolve critical situations or  mitigate risks, often in real-time. 
The effectiveness of this human oversight, highly depends on the human's situation awareness (SA)~\cite{endsleySA} (more broadly referred to as epistemic access~\cite{sterz}), which is rooted in the user's \emph{perception of information} displayed by the the monitoring interface~\cite{endsley2024situation}.
Missing information can lead to undetected errors which can have serious consequences, making it essential to develop and design monitoring interfaces that effectively support users perception of information~\cite{endsley2024situation}, as they have to keep track of numerous visual details and interface elements. 

In this regard, we see great potential for intelligent user interfaces (IUIs) to support human oversight of autonomous systems, in particular for real-time monitoring and time-critical decision making. Consider for example the case of a human overseeing a fleet of delivery drones. Visual highlights, auditory alarms or other dynamic UI elements can effectively draw attention to critical information, such as dysfunctional rotors or strong winds, helping users promptly address safety-critical situations~\cite{Li2020,wu2025understanding, wu2024enhancing}. However, simply highlighting every potentially critical event can lead to alarm floods and attentional overload~\cite{wan2024}, inappropriate reliance on alarms~\cite{wureleyance}, or alarm fatigue as a consequence of too many false warning~\cite{cvach2012monitor}. This problem is exacerbated by the fact that attentional capacity during monitoring varies substantially across individuals and situations, depending for example on current cognitive load~\cite{das24}, users’ expertise, or individual strategies~\cite{feit20}. Therefore, we see a need for intelligent monitoring interfaces that provide personalized oversight support, for example by adaptively prioritizing and scheduling alarms based on the user’s current knowledge or monitoring behavior.

In this paper, we explore RL-based UI adaptation as an approach to personalize monitoring support.   RL  has recently been proposed as a method to learn adaptive interaction policies directly from user behavior rather than relying on hand-crafted rules or heuristics~\cite{langerak24, Figueiredo24}. In our setting, the user, the monitored system, and the UI form the RL agent’s environment. It learns an optimal policy for deciding \emph{when} to alert a user and for \emph{how long}, based on the user’s current knowledge and the system's state.  By framing interface adaptation as a sequential decision problem, RL enables the discovery of personalized adaptation policies that balance the benefits of alerting against the cognitive cost of interruptions, potentially supporting more effective oversight over time.

A key challenge in developing RL-based adaptive interfaces is simulating the environment in which an optimal adaptation policy can be learned before real-world deployment. Here, the environment consists of the monitored drones and the user monitoring them, specifically their visual attention and perception of the drones' states which is affected by issued alarms. Learning via interaction with real users is infeasible due to time, cost, and safety constraints in such a setting. On the other hand, relying solely on pre-collected user data is highly limiting and does not capture the dynamics of ongoing user behavior~\cite{langerak24}. To address this, we propose leveraging \emph{models} of users' gaze behavior to generate diverse user data for training and simulate realistic attentional responses to the agent's alarm decisions.

In the following, we describe our first steps towards implementing such an RL-based approach for learning UI adaptation policies with user simulations. Building on our previous work, we explore this in the context of human oversight of delivery drones~\cite{wu2025understanding, wu2024enhancing, wureleyance}. We first define our use case and introduce gaze modeling. We then describe how we integrate gaze simulation with our RL architecture. Finally, we present preliminary qualitative findings showing that RL-based highlighting strategies can potentially outperform rule-based highlighting approaches. These initial results indicate that integrating human models with RL is a promising approach for supporting human oversight, which we are planning to empirically assess in the future.

\section{Use Case and Models of User Attention}
\label{sec:usecase}

We consider a multi-drone oversight scenario in which users must supervise the states of several drones in parallel \cite{wu2025understanding, wan2019identifying}. Following our previous work~\cite{wu2025understanding} the user oversees $N = 4$ drones using a dashboard-like interface (see \cref{sec:appendix_drone}). The drone information to supervise is given by a set of real-valued attributes $Attr$, displayed along with representative icons on the screen, e.g.,  containing horizontal velocity, battery level (as a number in $[0, 1]$, $0$ meaning depleted) or rotor functionality ($0$ for off, $1$ for on). In total, there are $|Attr| = 8$ attributes and
$N \cdot |Attr| = 32$ drone-attribute pairs. 

Maintaining situation awareness in this setting requires users to continuously monitor many interface elements while also being able to quickly notice and respond to critical events. In such settings, highlighting is commonly used to alarm users and direct their attention to critical information~\cite{wu2025understanding,wan2019identifying, veraDronePaper}. 
However, deciding when and where to guide user attention without overwhelming or disrupting the user with too many highlights, is a challenging design question.
We propose to address this challenge with an adaptive highlighting approach: by reasoning about users' visual attention, that is, which parts of the interface the user is likely to attend to at a given moment and how this attention evolves over time, we can dynamically decide at any given time whether a highlight can meaningfully support users in their monitoring activity. Since user attention depends (among other things) on the visual state of the interface and the interface’s adaptation decisions (e.g., highlighting), this reasoning can be based on \emph{models of visual attention}.

A promising and widely adopted approach for modeling visual attention in monitoring tasks is saliency prediction \cite{wu2025understanding}. 
Saliency models take a visual representation of the interface as input and predict user attention in the form of a saliency map. While most works in UI saliency prediction focus on static images \cite{jiang2023ueyes}, recent models of saliency prediction in natural scenes also incorporate temporal information, for example by processing sequences of past frames such as video clips \cite{min2019tased} or by predicting attention distributions over future time steps \cite{aydemir2023tempsal}. Temporal information is important to consider in this scenario, where the interface changes (e.g., displaying highlights, updating drone information). 

In this work, we use saliency models as a probabilistic user attention model that serves as the backbone of our adaptive interface. Specifically, we model how user attention changes in response to visual highlights using the temporal saliency model TASED-Net \cite{min2019tased}, which we fine-tuned to predict visual attention in dynamic monitoring interfaces in our previous work~\cite{wu2025understanding}. By embedding this saliency-based attention prediction within a RL framework, our interface can learn highlighting strategies that adapt over time based on the predicted attention dynamics and the resulting situation awareness of the user. We describe this in more detail next.
While we chose this specific model as a first step, our RL framework can easily integrate other alternative gaze models such as simple heuristic ones (e.g., assuming users attend to highlighted elements immediately) or more complex computational models that predict individual scan paths~\cite{jiang2024eyeformer}. Systematically exploring the impact of model fidelity on training efficiency and outcome will be part of our future work.

\section{Designing the RL Environment}

\newcommand{\sint}{s_{att}}
\newcommand{\susr}{s_{usr}}
\newcommand{\shlt}{s_{hlt}}
\newcommand{\reals}{\mathbb{R}}


To learn a UI highlighting policy to support drone oversight, we model the oversight environment (i.e., the interface displaying the drones' attributes, as well as the user's knowledge of these attributes) as a Markov decision process (MDP) and use standard off-the-shelf RL algorithms 
\cite{sutton1998reinforcement} to learn a user interface adaptation policy.
The MDP models the combined behavior of interface and user as a transition system.
States in that system represent drone attribute values, current user knowledge, and highlighted UI 
elements.
Actions consist in choosing which attributes to highlight. During training, the environment 
transitions to new states by simulating the drones' behavior (e.g. flight path, battery levels 
etc, see our previous work~\cite{wu2024enhancing, wu2025understanding}) and by stochastically 
simulating the user's visual attention on the interface using the gaze models described in \cref{sec:usecase}.
In other words, the MDP models the interface as an agent which takes  highlighting actions that 
affect its environment, that is the user's belief about the drones' current attributes, simulated 
via models of human visual attention. 
As the specific RL algorithm, we use proximal policy optimization (PPO, \cite{schulman2017proximal}).
In what follows, we introduce the MDP in more detail.



\paragraph{State and Action Representation}
We represent a state $s = (\sint, \susr, \shlt)$ of the environment using three different 
components:
The attribute state $\sint$, the user state $\susr$ and the highlight state $\shlt$.
The attribute state component $\sint$ represents the current state of the attributes, i.e., the 
values of each drone-related icon visible on the screen.
Given a fixed number of drones $N$ and a fixed set of attributes $Attr$ to oversee for each drone,
$\sint$ is an assignment $\sint : \{ 1, \dots N \} \times Attr \to \reals$ that specifies the 
currently displayed value for each drone-attribute pair.
The user state $\susr : \{ 1, \dots N \} \times Attr \to \reals$, on the other hand, represents the current user \emph{belief} about what the 
attribute values are.
Lastly, the highlight state is a binary vector $\shlt : \{ 1, \dots N \} \times Attr \to \{ 0, 1\}$
that assigns a drone-attribute pair the value $1$ if and only if it is currently highlighted by
the user interface.

An action $a$ in the environment is a choice of highlights, and thus takes the same form as the
highlight state, i.e., each action $a$ is a function $a : \{ 1, \dots N \} \times Attr \to \{ 0, 1\}$.


\paragraph{State Transition Function}

We assume that initially the environment starts in a state that the user is acquainted with,
so that $\sint = \susr$.
Initially, nothing is highlighted, i.e., $\shlt \equiv 0$.

Given any current state $s$ and highlighting decision $a$ 
our MDP-environment simulator stochastically transitions to the next state as follows:
\begin{enumerate}
    \item First, the attribute state $\sint$ is advanced by a fixed time interval, simulating the changes in the drones's attributes, which are independent from the rest of the state.
    \item Secondly, we set the highlight state of the interface to match the 
          action of the agent, i.e., $\shlt \gets a$.
    \item Next, the user's gaze behaviour is simulated using the visual attention model 
          described in the previous section, which takes an image of the current interface as input 
          (reflecting the updated attributes $\sint$ and any highlighted icons $\shlt$) and outputs
          a probability distribution $P$ over the drone-attribute pairs $\{1, \dots, n\} \times Attr$ at
          which the user may look next.
          We sample one of the drone-attribute pairs $(d, \alpha) \sim P$, committing to this one as the 
          one that the user has looked at.
          Consider, that if an icon is highlighted, the saliency model will predict a higher likelihood for 
          this icon to be looked at.
          If nothing is highlighted, the fine-tuned model gives higher weight to icons real users have  
          frequently been looking at during the drone monitoring task (see~\cite{wu2025understanding}).
          We update the user's belief state by setting $\susr(I) \gets \sint(I)$, leaving all other 
          attributes in the user belief at their previous values. The correctness of the user's belief 
          about not-looked-at attributes thus degrades over time as the drone attribute values change.
\end{enumerate}


\paragraph{Rewards}

The reward in our environment consists of two components. First, we penalize inaccuracies in user's beliefs through a weighted sum over differences between user knowledge (i.e., $\susr$) and 
actual attribute state (i.e, $\sint$), i.e., 
\begin{equation}
d(\sint, \susr)
\coloneq
\sum_{1 \leq d \leq N} \sum_{\alpha \in Attr}
w(\alpha) \cdot |\sint(d, \alpha) - \susr(d, \alpha)|.
\label{eq:distance}
\end{equation}
The weighting function  $w : Attr \to \reals^+$ associates each attribute with a weight representing its 
monitoring importance relative to other attributes (e.g., the rotor off/on attribute is given a weight of 
$100$, the distance to the drone's target only a $1$).

Second, we give a fixed penalty $H \in \reals_0^+$ for each highlight displayed on the screen, to reflect that highlighting has a cost (e.g., it consumes user attention which cannot be directed to other information~\cite{wu2025understanding}, might contribute to alarm fatigue~\cite{cvach2012monitor}, etc.).

The overall reward given for a state $s$ is the sum of both penalties, i.e., 
\begin{equation*}
R(s) \coloneq -d(\sint, \susr) - H * (\sum_{1 \leq d \leq N} \sum_{\alpha \in Attr} \shlt(d,\alpha))
\end{equation*}
By maximizing this reward function, the RL agent should learn to highlight the most important attributes that the user has the least knowledge about, while using highlighting as sparsely as possible.
The hyperparameters, including the attribute weights and highlight penalty will largely influence the learned policy and will need to be tuned to the specific application. Our future work will explore how to better ground them in the literature, such as  explicitly capturing factors that influence the cognitive cost of highlighting, and considering not only users current knowledge ($\susr$, i.e, Level 1 SA
(see~\cite{endsleySA})) but users' ability to predict the state into the future (i.e., Level 3 SA
(see~\cite{endsleySA})).

\section{Training and Results}

The training scenario we designed contains multiple occurrences of different critical 
situations in close succession, in which important traits like rotor functionality of a drone 
or its stay in a no-fly zone must be recognized quickly by the supervisor.
We trained an interface policy for the drone supervision scenario introduced in
\cref{sec:usecase} with a highlight penalty of $H = 500$.
This is high enough such that the immediate penalty given for each drone-attribute 
pair, i.e., the weighted distance it contributes to \cref{eq:distance}, is dominated it.
Therefore, the interface agent must foresee whether displaying a
highlight is beneficial in the long run, as future penalties will be avoided when the 
user's knowledge is updated.

For the hyperparameters of PPO, we use mostly standard settings with minor adjustments
(cf., \cite{andrychowicz2020matters}).
All detailed settings are listed in the appendix, along with reports of the recorded
training metrics.

\begin{figure}[t]
    \begin{subfigure}[t]{.315\textwidth}
        \includegraphics[width=\linewidth]{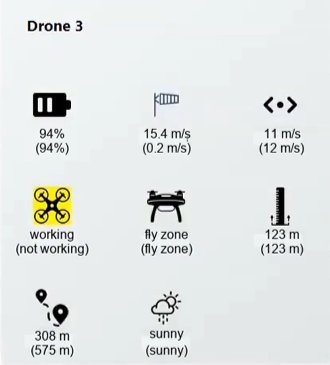}
        \caption{
        State of the scenario as drone 3 experiences high wind speeds
        and recovers rotor functionality.
        The interface policy decides to highlight the rotor icon.
        }
    \end{subfigure}
    \hfill
    \begin{subfigure}[t]{.315\textwidth}
        \includegraphics[width=\linewidth]{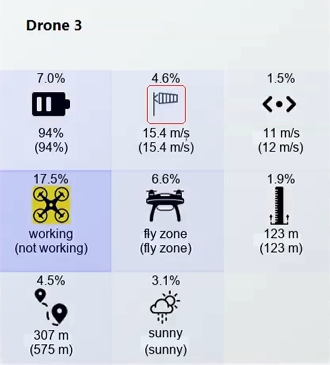}
        \caption{
        The simulator samples from the saliency model (probabilities shown in blue with values above each icon) and simulates the next fixation of the user (red) to fall on the wind speed icon.
        }
    \end{subfigure}
    \hfill
    \begin{subfigure}[t]{.315\textwidth}
        \includegraphics[width=\linewidth]{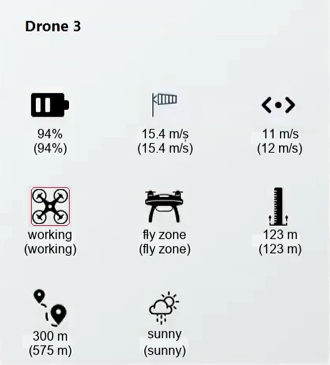}
        \caption{
        In the next time step, the user is simulated to fixate the highlighted
        rotor icon, at which point the interface policy decides to remove all highlights from the interface.
        }
    \end{subfigure}
    \centering
    \caption{Example behavior of the learned interface policy as a critical situation 
    affecting drone 3 occurs in the scenario.
    Only the region of the interface dedicated to drone 3 is shown.
    Each icon is annotated with its actual displayed value below, as well as the value of the user 
    knowledge state in parentheses further below.
    }
    \label{fig:scenario_examples}
\end{figure}


Our preliminary qualitative assessment indicates that the policy is often able to successfully 
highlight critical situations (e.g., when a drone rotor fails) or when a stark contrast 
between user knowledge and actual drone state exists.
An example situation where this can be observed is depicted in 
\cref{fig:scenario_examples}.
This example depicts a critical situation in which one of the four drones experiences a dangerous wind speed,
while at the same time recovering from a malfunctioning rotor.
The policy decided to highlight the recovered rotor.
The simulated user first looks at the both wind speed (maybe due to inaccuracies in gaze movements, or a delay in noticing the actual highlight) and only then at the highlighted icons. As the user is then aware of both situations, the policy removes all highlights and also ignores the still critical windspeed. A typical rule-based highlighting approach (e.g. highlight every critical situation for five seconds) might have unnecessarily highlighted the wind speed even though the user is already aware of it, or might have led the user to miss the critical information about the recovered rotor by drawing too much attention on the wind speed~\cite{wu2025understanding}.

However, our we also saw instances where our preliminary interface  agent failed to recognize such scenarios and chose not to display a highlight, although doing so would most likely be beneficial.
In general, the behavior of the policy is heavily influenced by the gaze simulation and highlight penalty.
If the gaze simulation predicts that the user will not look at the highlighted icon with high enough probability, or the highlight penalty is too high, then the policy is discouraged from highlighting this icon.
The opposite effect can of course also be problematic, as it would lead to overly 
opportunistic policies, distracting the user more than aiding him in his supervision.
A major focus of our ongoing work is thus to systematically explore how sensitive the interface training is to the fidelity of the gaze simulation model and empirically evaluate the influence of different highlighting policies in actual monitoring scenarios with real users.


\section{Conclusion}
This paper presents an initial step toward RL-based adaptive oversight interfaces that make use of models of human visual attention to learn personalized and dynamic strategies for alarming oversight persons about critical information or potential problems in their situation awareness. While our preliminary results suggest that combining RL with gaze-based user simulations is a promising direction, substantial challenges remain before such approaches can be deployed in real-world oversight settings. Our key next steps include empirically evaluating adaptive alarm strategies with real users, systematically examining how the fidelity of gaze models influences learned policies, and analyzing computational trade-offs between model realism and training efficiency. More complex oversight scenarios, such as alarm floods or concurrent critical events, are also needed to stress-test learned policies where conventional highlighting strategies are known to fail. In addition, reward design remains a central challenge~\cite{Figueiredo24}: beyond aligning user knowledge with system state, future reward formulations should also capture cognitive and psychological costs to mitigate alarm fatigue and information overload. At a higher level, future work should investigate how intelligent interfaces might affect the effectiveness of human oversight~\cite{sterz} in the long-term, including issues such as over- or under-reliance, deskilling, or automation bias. We hope these directions resonate with workshop participants and stimulate discussion on how intelligent user interfaces can support the evolving challenges of human oversight.

%
\begin{acknowledgments}
This work was funded by the Deutsche Forschungsgemeinschaft (DFG, German Research Foundation) – project number 389792660 – TRR 248 (see \url{https://perspicuous-computing.science}).
\end{acknowledgments}

\bibliography{biblio}

\clearpage

\begin{appendix}

\crefalias{section}{appendix}

\section{Drone Monitoring Interface}
\label{sec:appendix_drone}
\begin{figure}[htbp]
  \centering
\includegraphics[width=0.75\linewidth]{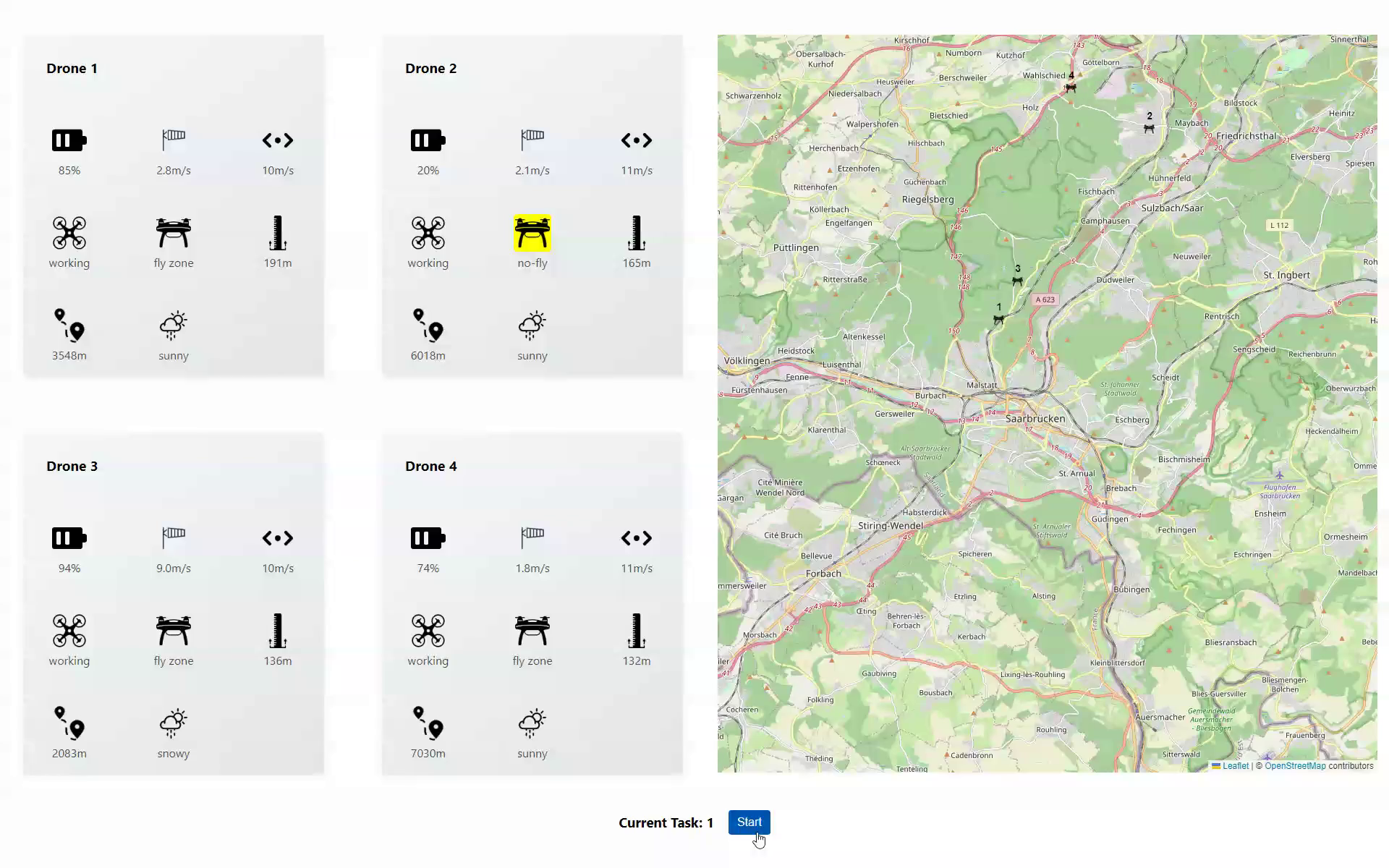}
  \caption{Dashboard-like interface for supervising four drones. Each drone panel displays eight attribute icons, while a map view provides spatial context. The interface is adopted from our previous work~\cite{wu2025understanding}. The UI agent learns to control the highlighting of icons. The map on the right is only for increasing the realism of the interface during user studies and is disregarded by the UI agent.}
  \label{fig:ui_overview}
\end{figure}

Figure~\ref{fig:ui_overview} shows the dashboard-like multi-drone monitoring interface used in our oversight scenario. The interface arranges drone-specific information in a set of dedicated panels on the left side, while a map view on the right provides spatial context by visualizing the current positions of all drones. Each panel contains a fixed set of icons representing the drone’s state, enabling users to monitor multiple drones in parallel. Critical situations are visually indicated by color-based highlighting of individual icons, while the overall layout and iconography remain fixed over time so that changes in visual saliency are driven solely by dynamic highlights and updated attribute values.

\section{Training Parameters \& Metrics}
We used the following training parameters for the training procedure.

\begin{itemize}
    \item Discount factor: $\gamma=0.95$
    \item Generalized advantage estimation parameter: $\lambda=0.95$
    \item Clip epsilon: $\epsilon = 0.2$
    \item Total training samples: $20.000.000$
    \item Training batch size: $16384$
    \item Training epochs for each batch of data: $10$
    \item Mini-batch size: $256$
\end{itemize}

We employ Adam \cite{adam2014method} with a learning rate of $3 \cdot 10^{-4}$ 
and $(\beta_1, \beta_2) = (0.9, 0.999)$ as the optimizer 
and linearly decrease the learning rate towards zero after each training episode.

The critic and policy models are multi-layer perceptrons with two layers of width $64$.

\begin{figure}
    \centering
    \includegraphics[height=.25\textheight]{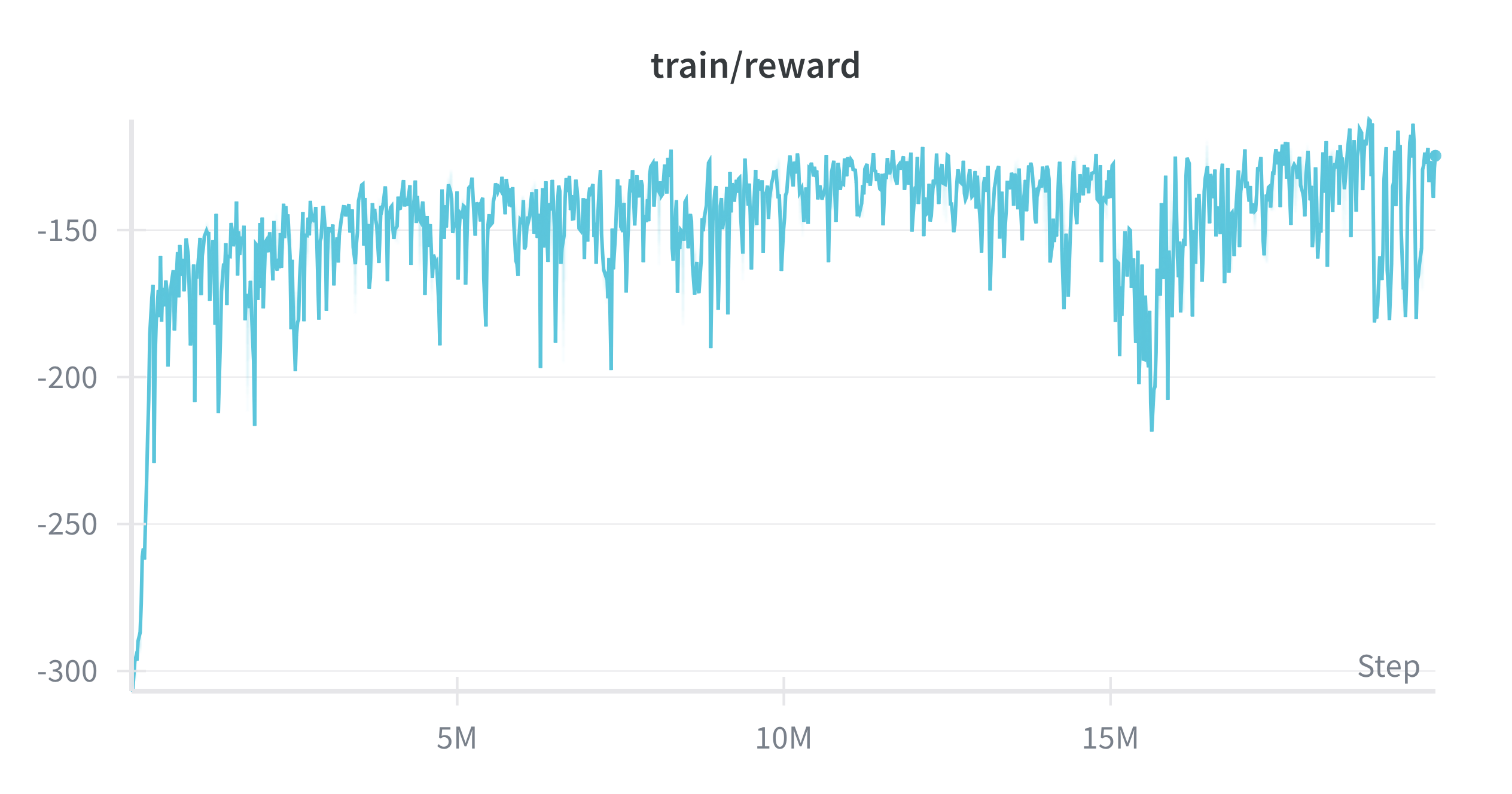}
    \caption{
    Training curve of PPO with environment parameter $H = 500$ for the highlight penalty.
    On the x-axis, the number of training samples is shown.
    On the y-axis, the mean total reward per collected episode is shown.
    }
    \label{fig:training_curve}
\end{figure}

The training curve in terms of the training reward over time is depicted in 
\cref{fig:training_curve}.

\end{appendix}

\end{document}